\documentclass{article}
\usepackage{amssymb}
\usepackage{amsmath}
\usepackage{amsfonts}
\usepackage{graphicx}

\usepackage{eusipco}                       

\title{The discrete Fourier transform:\\ A canonical basis of eigenfunctions}

\name {Shamgar Gurevich${}^1$, Ronny Hadani${}^2$, and Nir
Sochen${}^3$}

\address{ ${}^1$Department of Mathematics, University
of California, \\
Berkeley, CA 94720, USA \\
E-mail: shamgar@math.berkeley.edu \\
\\
${}^2$Department of Mathematics, University of Chicago\\
, IL 60637, USA\\
E-mail: hadani@math.uchicago.edu \\
\\
${}^3$School of Mathematical Sciences, Tel Aviv University\\
Tel Aviv 69978, Israel\\
E-mail: sochen@math.tau.ac.il}

\begin{document}

\maketitle

\begin{abstract}
The discrete Fourier transform (DFT) is an important operator which 
acts on the Hilbert space of complex valued functions on the ring $\mathbb{Z}/N\mathbb{Z}$. 
In the case where $N=p$ is an odd prime number, we exhibit a canonical basis $\Phi$ of eigenvectors 
for the DFT. The transition matrix $\Theta$ from the standard basis to 
$\Phi$ defines a novel transform which we call the \textit{discrete oscillator transform} (DOT for short).
Finally, we describe a fast algorithm for computing $\Theta$ in certain cases.
\end{abstract}


\section{Introduction}
The discrete Fourier transform (DFT) is probably one of the most
important operators in modern science. It is omnipresent in
various fields of discrete mathematics and engineering, including
combinatorics, number theory, computer science and, last but
probably not least, digital signal processing. Formally, the DFT
is a family $\left \{ F_{N}\right \} $ of \
unitary operators, where each $F_{N}$ acts on the Hilbert space $\mathcal{H}%
_{N}\mathcal{=%
\mathbb{C}
}\left(
\mathbb{Z}
/N%
\mathbb{Z}
\right) $ by the formula

\begin{equation*}
F_{N}\left[ f\right] \left( w\right) =\frac{1}{\sqrt{N}}\sum
\limits_{t\in
\mathbb{Z}
/N%
\mathbb{Z}
}e^{\frac{2\pi i}{N}wt}f\left( t\right) .
\end{equation*}

Although, so widely used, the spectral properties of the DFT
remains to some extent still mysterious. For example, the
calculation of the multiplicities of its eigenvalues, which was
first carried out by Gauss, is quite involved and requires a
multiple of number theoretic manipulations \cite{AT}.

A primary motivation for studying the eigenvectors of the DFT
comes from digital signal processing. Here, a function is
considered in two basic realizations: The time realization and the
frequency realization. Each realization, yields information on
different attributes of the function. The
DFT operator acts as a dictionary between these two realizations%

\begin{equation*}
\text{\textbf{Time} }\overset{F_{N}}{\rightleftarrows }\text{ \textbf{%
Frequency.} }
\end{equation*}

From this point of view, it is natural to look for a
diagonalization basis, namely, a basis of eigenvectors (eigen
modes) for $F_{N}$. In this regard, the main conceptual difficulty
comes from the fact that the diagonalization
problem is ill-defined, since, $F_{N}$ is an operator of order 4, i.e., $%
F_{N}^{4}=Id$, which means that it has at most four eigenvalues
$\pm 1$,$\pm i$, therefore each appears with large multiplicity
(We assume $N\gg 4$).

An interesting approach to the resolution of this difficulty,
motivated from results in continuous Fourier analysis, was developed by Gr\"{u}nbaum in \cite%
{G}. In that approach, a tridiagonal operator $S_{N}$ which commutes with $%
F_{N}$ and admits a simple spectrum is introduced. This enable him
to give a basis of eigenfunctions for the DFT. Specifically,
$S_{N}$ appears as a certain discrete analogue of the differential
operator $D=\partial _{t}^{2}-t^{2}$ which commutes with the
continuous Fourier transform.

\subsection{Main results of this paper}

In this paper we describe a representation theoretic approach to
the diagonalization problem of the DFT in the case when $N=p$ is
an odd prime number. Our approach, puts to the forefront the Weil
representation \cite{W} of the finite symplectic group
$Sp=SL_{2}\left( \mathbb{F}_{p}\right) $ as the fundamental object
underlying harmonic analysis in the finite setting. Specifically,
we exhibit a canonical basis $\Phi _{p}$ of eigenvectors for the
DFT. We also describe the transition matrix $\Theta _{p}$ from the
standard basis to $\Phi _{p}$, which we call the \textit{discrete
oscillator transform} (DOT for short). In addition, in the case
$p\equiv 1 ($mod $4)$, we describe a fast
algorithm for computing $\Theta _{p}$ (FOT for short).

It is our general feeling that the Weil representation yields a
transparent explanation to many classical results in finite
harmonic analysis. To justify this claim, we describe an
alternative method for calculating the multiplicities of the
eigenvalues for the DFT, a method we believe is more suggestive
then the classical calculations.

The rest of the introduction is devoted to a more detailed account
of the main ideas and results of this paper.

\subsection{Symmetries of the DFT}

Let us fix an odd prime number $p$ and for the rest of the
introduction suppress the subscript $p$ from all notations.
Generally, when a (diagonalizable) linear operator $A$ has
eigenvalues
admitting large multiplicities, it suggest that there exists a group $%
G=G_{A}\subset GL\left( \mathcal{H}\right) $ of "hidden"
symmetries consisting of operators which commute with $A$. \ Alas,
usually the problem of computing the group $G$ is formidable and,
in fact, equivalent to the problem of diagonalizing $A$. If the
operator $A$ \ arise "naturally", there is a chance that the group
$G$ can be effectively described. In preferred situations, $G$ is
commutative and large enough so that all degeneracies are resolved
and the spaces of common eigenvectors with respect to $G$ are
one-dimensional. The basis of common eigenvectors with respect to
$G$ establishes a distinguish choice of eigenvectors for $A$.
Philosophically, we can say that it \ is more correct to consider
from start the group $G$ instead of the single operator $A$.

Interestingly, the DFT operator $F=F_{p}$ admits a natural group
of symmetries $G_{F}$, which, in addition, can be effectively
described using the Weil representation. For the sake of the
introduction, it is enough to know that the Weil representation in
this setting is a unitary representation $\rho :Sp\rightarrow
U\left( \mathcal{H}\right) $ and the key observation is that $F$
is proportional to a single operator $\rho \left(
\mathrm{w}\right) .$ The group $G_{F}$ is the image under $\rho $
of the centralizer subgroup $T_{\mathrm{w}}$ of \textrm{w} in
$Sp.$

\subsection{The algebraic torus associated to the DFT}

The subgroup $T_{\mathrm{w}}$ can be computed explicitly and is of
a very "nice" type, it consists of rational points of a maximal algebraic torus in $%
Sp$ which in plain language means that it is maximal commutative
subgroup in $Sp$, consisting of elements which are diagonalizable
over some field
extension. Restricting the Weil representation to the subgroup $T_{\mathrm{w}%
}$ yields a collection $G_{F}=\left \{ \rho \left( g\right) :g\in T_{\mathrm{%
w}}\right \} $ of commuting operators, each acts unitarily on the
Hilbert space $\mathcal{H}$ and commutes with $F$. This, in turn,
yields a decomposition, stable under Fourier transform, into
character spaces

\begin{equation}
\mathcal{H=}\bigoplus \mathcal{H}_{\chi },  \label{dec1_eq}
\end{equation}%
where $\chi $ runs in the set of (complex valued) characters of $T_{\mathrm{w%
}}$, namely, if $v\in \mathcal{H}_{\chi }$ then $\rho \left(
g\right) v=\chi
\left( g\right) v$. The main technical statement of this paper, Theorem 3, 
roughly says that $\dim \mathcal{H}_{\chi }=1$ for
every $\chi $ which appears in (\ref{dec1_eq}). Choosing a unit
representative $\phi _{\chi }\in \mathcal{H}_{\chi }$ for every
$\chi $, gives the canonical basis $\Phi =\left \{ \phi _{\chi
}\right \} $ of eigenvectors for $F$. The oscillator transform
$\Theta $ sends a function $f\in \mathcal{H}$ to the coefficients
in the unique expansion
\begin{equation*}
f=\sum a_{\chi }\phi _{\chi }.
\end{equation*}
The fine behavior of $F$ and $\Theta $ is governed by the\ (split
type) structure of $T_{\mathrm{w}}$, which changes depending on
the value of the prime $p$ modulo $4$. This phenomena has several
consequences. In particular, it gives a transparent explanation to
the precise way the multiplicities of the eigenvalues of $F$
depend on the prime $p$. Another, algorithmic, consequence is
related to the existence of a fast algorithm for computing $\Theta
$.

\subsection{Properties of eigenvectors}

The character vectors $\phi _{\chi }$ satisfy many interesting
properties and are objects of study in their own right. A
comprehensive treatment of this aspect of the theory appears in
\cite{GHS}.

\subsection{Generalizations}

\subsubsection{Field extensions}

All the results in this paper were stated for the basic finite field $%
\mathbb{F}_{p}$, for the reason of making the terminology more
accessible. In fact, all the results can be stated and proved for
any field extension of the form $\mathbb{F}_{q}$, $q=p^{n}$. In most places one
should only need to replace $p$ by $q$.

\subsection{Structure of the paper}
We begin by discussing the finite Heisenberg group and the
Heisenberg representation. Next we introduce the Weil
representation of the finite symplectic group, first it is
described in abstract terms and then more explicitly invoking the
idea of invariant presentation of an operator. We proceed to
discuss the theory of tori in the one-dimensional Weil
representation, we explain how to associate to a maximal torus
$T\subset SL_{2}$, a transform $\Theta _{T}$ called the oscillator
transform. We describe a fast algorithm for computing $\Theta
_{T}$ in the case $T$ is a split torus. The theory is then applied
to the specific torus associated with the DFT operator. We finish
with a treatment of the multiplicity problem for the DFT, from the
representation theoretic perspective.

\subsection{Acknowledgements}

It is a pleasure to thank J. Bernstein for his
interest and guidance. We acknowledge R. Howe for sharing with us
some unpublished results of his. We appreciate several discussions we had with A. Gr\"{u}%
nbaum, W. Kahan, B. Parlett, B. Porat, A. Sahai on the DFT. We
thank A. Weinstein, J. Wolf and M. Zworski for interesting
conversations related to the Weil representation. Finally,
we thank P. Diaconis, K. Ribet, B. Poonen, M. Haiman and M. Gu for the
opportunities to present this work in the MSRI, number theory,
representation theory and scientific computing seminars at
Berkeley during February 2008.

\section{The Oscillator Transform\label{pre_sec}}

\subsection{The Heisenberg group}

Let $(V,\omega )$ be a two-dimensional symplectic vector space
over the finite field $\mathbb{F}_{p}$. The reader should think of $V$ as $\mathbb{F}%
_{p}\times \mathbb{F}_{p}$ with the standard form $\omega \left(
\left( \tau ,w\right) ,\left( \tau ^{\prime },w^{\prime }\right)
\right) =\tau w^{\prime }-w\tau ^{\prime }$. Considering $V$ as an
abelian group, it admits a non-trivial central extension called
the \textit{Heisenberg }group. \
Concretely, the group $H$ can be presented as the set $H=V\times \mathbb{F}%
_{p}$ with the multiplication given by%
\begin{equation*}
(v,z)\cdot (v^{\prime },z^{\prime })=(v+v^{\prime },z+z^{\prime }+\tfrac{1}{2%
}\omega (v,v^{\prime })).
\end{equation*}

The center of $H$ is $\ Z=Z(H)=\left \{ (0,z):\text{ }z\in \mathbb{F}%
_{p}\right \} .$ The symplectic group $Sp=Sp(V,\omega )$, which in
this case is isomorphic to $SL_{2}\left( \mathbb{F}_{p}\right) $,
acts by automorphism of $H$ through its action on the
$V$-coordinate.

\subsection{The Heisenberg representation\label{HR}}

One of the most important attributes of the group $H$ is that it
admits, principally, a unique irreducible representation. The
precise statement goes as follows. Let $\psi :Z\rightarrow
\mathbb{C}
^{\times }$ be a character of the center. For example we can take
$\psi \left( z\right) =e^{\frac{2\pi i}{p}z}$.

\noindent {\bf Theorem 1.}(Stone-von Neumann)

\label{S-vN}{\em There exists a unique (up to isomorphism)
irreducible unitary representation $(\pi ,H,\mathcal{H)}$ with the
center acting by $\psi ,$ i.e., $\pi _{|Z}=\psi \cdot
Id_{\mathcal{H}}$.}


The representation $\pi $ which appears in the above theorem will
be called the \textit{Heisenberg representation}.

\subsubsection{Standard realization of the Heisenberg representation\label%
{standard_subsub}.}

The Heisenberg representation $(\pi ,H,\mathcal{H)}$ can be
realized as follows: $\mathcal{H}$ is the Hilbert space $%
\mathbb{C}
(\mathbb{F}_{p})$ of complex valued functions on the finite line,
with the standard Hermitian product. The action $\pi $ is given by
$\pi (\tau ,0)[f]\left( t\right) =f\left( t+\tau \right) $, $\pi
(0,w)[f]\left( t\right) =\psi \left( wt\right) f\left( t\right) $
and $\pi (z)[f]\left( t\right) =\psi \left(
z\right) f\left( t\right) $. We call this explicit realization the \textit{%
standard realization}.

\subsection{The Weil representation\label{Wrep_sub}}

A direct consequence of Theorem 1 
is the existence of a projective representation $\widetilde{\rho
}:Sp\rightarrow PGL(\mathcal{H)}$. The
construction of $\widetilde{\rho }$ out of the Heisenberg representation $%
\pi $ is due to Weil \cite{W} and it goes as follows. Considering
the Heisenberg representation $\pi $ and an element $g\in Sp$, one
can define a new representation $\pi ^{g}$ acting on the same
Hilbert space via $\pi
^{g}\left( h\right) =\pi \left( g\left( h\right) \right) $. Clearly both $%
\pi $ and $\pi ^{g}$ have the same central character $\psi $ hence
by Theorem 1 
they are isomorphic. Since the space $\mathsf{Hom}%
_{H}(\pi ,\pi ^{g})$ is one-dimensional, choosing for every $g\in
Sp$ a non-zero representative $\widetilde{\rho }(g)\in
\mathsf{Hom}_{H}(\pi ,\pi ^{g})$ gives the required projective
representation. In more concrete terms, the projective
representation $\widetilde{\rho }$ is characterized by the formula

\begin{equation}
\widetilde{\rho } \left( g\right) \pi \left( h\right) \widetilde{\rho }\left( g^{-1}\right)
=\pi \left( g\left( h\right) \right) ,  \label{Egorov}
\end{equation}
for every $g\in Sp$ and $h\in H$. It is a peculiar phenomenon of
the finite field setting that the projective representation
$\widetilde{\rho }$ can be linearized into an honest
representation.

\noindent {\bf Theorem 2}
\label{linearization} {\em There exists a unique\footnote{%
Unique, except in the case the finite field is $\mathbb{F}_{3}$.
For the canonical choice in the latter case see \cite{GH1}.}
linear representation
\begin{equation*}
\rho :Sp\longrightarrow GL(\mathcal{H)},
\end{equation*}
which satisfies equation (\ref{Egorov}).}


\subsubsection{Invariant presentation of the Weil representation}

Let us denote by $%
\mathbb{C}
\left( H,\psi \right) $ the space of (complex valued) functions on
$H$ which are $\psi $-equivariant with respect to the action of
the center, namely, a function $f\in
\mathbb{C}
\left( H,\psi \right) $ satisfies $f\left( zh\right) =\psi \left(
z\right) f\left( h\right) $ for every $z\in Z$, $h\in H$. Given an
operator $A\in \mathsf{End}\left( \mathcal{H}\right) $, it can be
written in a unique way as $A=\pi \left( K_{A}\right) $, where
$K_{A}\in
\mathbb{C}
\left( H,\psi ^{-1}\right) $ and $\pi $ denotes the extended
action $\pi \left( K_{A}\right) =\sum \limits_{h\in H}K_{A}\left(
h\right) \pi \left( h\right) .$ The function $K_{A}$ is called the
\textit{kernel of }$A$ and it
is given by the \textit{matrix coefficient }%
\begin{equation}
K_{A}\left( h\right) =\frac{1}{\dim \mathcal{H}}Tr\left( A\pi
\left( h^{-1}\right) \right) .  \label{Weyl_eq}
\end{equation}

In the context of the Heisenberg representation, formula
(\ref{Weyl_eq}) is usually referred to as the \textit{Weyl
transform}. Using the Weyl transform one is able to give an
explicit description of the Weil representation. The idea
\cite{GH1} is to write each operator $\rho \left( g\right) $,
$g\in Sp$ in terms of its kernel function $K_{g}=K_{\rho \left(
g\right) }\in
\mathbb{C}
\left( H,\psi ^{-1}\right) $. The following formula is taken from \cite{GH1}%
\begin{equation}
K_{g}\left( v,z\right) =\frac{\sigma \left( -1\right) }{\dim \mathcal{H}}%
\sigma \left( \det \left( \kappa \left( g\right) +I\right) \right)
\psi \left( \tfrac{1}{4}\omega \left( \kappa \left( g\right)
v,v\right) +z\right) \label{inv_formula}
\end{equation}%
for every $g\in Sp$ such that $g-I$ is invertible, where $\sigma $
denotes the unique quadratic character (Legendre character) of the
multiplicative group $\mathbb{F}_{p}^{\times }$ and $\kappa $ is the Cayley transform $%
\kappa \left( g\right) =\frac{g+I}{g-I}$, $g\in Sp$.

\subsection{The theory of tori}
A maximal (algebraic) torus in $Sp$ is a maximal commutative
subgroup which becomes diagonalizable over some field extension.
There exists two conjugacy classes of maximal (algebraic) tori in
$Sp$. The first class consists of those tori which are
diagonalizable already over $\mathbb{F}_{p}$ or equivalently those
are the tori that are conjugated to the standard diagonal
torus%
\begin{equation*}
A=\left \{
\begin{pmatrix}
a & 0 \\
0 & a^{-1}%
\end{pmatrix}%
:a\in \mathbb{F}_{p}\right \} .
\end{equation*}
A torus in this class is called a \textit{split} torus. The second
class consists of those tori which become diagonalizable over a
quadratic extension $\mathbb{F}_{p^{2}}$ or equivalently those are
tori which are not
conjugated to $A.$ A torus in this class is called a \textit{non-split }%
torus (sometimes it is called inert torus)$.$

\noindent {\bf Example 1}(Example of a non-split torus)

{\em It might be suggestive to explain further the notion of
non-split torus by exploring, first, the analogue notion in the
more familiar setting of the
field $%
\mathbb{R}
$. Here, the standard example of a maximal non-split torus is the
circle
group $SO(2)\subset SL_{2}(%
\mathbb{R}
)$. Indeed, it is a maximal commutative subgroup which becomes
diagonalizable when considered over the extension field $%
\mathbb{C}
$ of complex numbers. The above analogy suggests a way to
construct an example of a maximal non-split torus in the finite
field setting as well.

Let us identify the symplectic plane $V=\mathbb{F}_{p}\times
\mathbb{F}_{p}$ with the quadratic extension $\mathbb{F}_{p^{2}}$.
Under this
identification, $\mathbb{F}_{p^{2}}$ acts on $V$ and for every $g\in $ $%
\mathbb{F}_{p^{2}}$ we have $\omega \left( gu,gv\right) =\det
\left( g\right) \omega \left( u,v\right) $, which implies that the
group
\begin{equation*}
T_{ns}=\left \{ g\in \mathbb{F}_{p^{2}}^{\times }:\det \left(
g\right) =1\right \}
\end{equation*}%
naturally lies in $Sp$. The group $T_{ns}$ is an example of a
non-split torus which the reader might think of as the\ "finite
circle". }


\subsubsection{Decompositions with respect to a maximal torus}

Restricting the Weil representation to a maximal torus $T\subset
Sp$ yields a decomposition

\begin{equation}
\mathcal{H=}\bigoplus_{\chi }\mathcal{H}_{\chi },
\label{decomp_eq}
\end{equation}%
where $\chi $ runs in the set $T^{\vee }$ of complex valued
characters of
the torus $T$. More concretely, choosing a generator\footnote{%
A maximal torus $T$ in $SL_{2}\left( \mathbb{F}_{p}\right) $ is a
cyclic
group, thus there exists a generator.} $t\in T$, the decomposition (\ref%
{decomp_eq}) naturally corresponds to the eigenspaces of the
linear operator $\rho \left( t\right) $. The decomposition
(\ref{decomp_eq}) depends on the split type of $T$. Let $\sigma
_{T}$ denote the unique quadratic character of $T$.

\noindent {\bf Theorem 3}  (\cite{GH2})\label{dec_thm} {\em If $T$
is a split torus, then }
\begin{equation*}
\dim {\cal{H}}_{\chi }=\left \{
\begin{array}{cc}
1 & \chi \neq \sigma _{T}, \\
2 & \chi =\sigma _{T}.%
\end{array}%
\right.
\end{equation*}%
{\em If $T$ is a non-split torus, then}
\begin{equation*}
\dim {\cal{H}}_{\chi }=\left \{
\begin{array}{cc}
1 & \chi \neq \sigma _{T}, \\
0 & \chi =\sigma _{T}.%
\end{array}%
\right.
\end{equation*}


\subsection{ The discrete oscillator transform associated to a maximal torus}

Let us fix a maximal torus $T$. Every vector $v\in \mathcal{H}$
can be written uniquely as a direct sum $v=\sum v_{\chi }$ with
$v_{\chi }\in
\mathcal{H}_{\chi }$ and $\chi $ runs in $I=\mathsf{Spec}_{T}\left( \mathcal{%
H}\right) $ - the spectral support of $\mathcal{H}$ with respect
to $T$
consisting of all characters $\chi \in T^{\vee }$ such that $\dim \mathcal{H}%
_{\chi }\neq 0$. Let us choose, in addition, a collection of unit vectors $%
\phi _{\chi }\in \mathcal{H}_{\chi }$, $\chi \in I$ and let $\phi
=$ $\sum
\phi _{\chi }$. We define the transform $\Theta _{T}=\Theta _{T,\phi }:%
\mathcal{H\rightarrow
\mathbb{C}
}\left( I\right) $ by $\Theta _{T}\left[ v\right] \left( \chi
\right) =\left
\langle v,\phi _{\chi }\right \rangle $. We will call the transform $%
\Theta _{T}$ the \textit{discrete} \textit{oscillator transform}
(DOT for short) with respect to the torus $T$ and the test vector
$\phi $.

\noindent {\bf Remark 1}
We note that in the case $T$ is a non-split torus, $\Theta _{T}$ maps $%
\mathcal{H}$ isomorphically to $\mathcal{%
\mathbb{C}
}\left( I\right) $. In the case $T$ is a split torus, $\Theta
_{T}$ has a kernel consisting of $v\in \mathcal{H}$ such that
$\left \langle v,\phi _{\sigma _{T}}\right \rangle =0$.

\subsubsection{The oscillator transform (integral form)}

Let $\mathcal{M}_{T}:%
\mathbb{C}
(T)\rightarrow
\mathbb{C}
(T^{\vee })$ denote the Mellin transform
\begin{equation*}
\mathcal{M}_{T}\left[ f\right] \left( \chi \right)
=\frac{1}{\#T}\sum \limits_{g\in T}\overline{\chi }\left( g\right)
f\left( g\right) ,
\end{equation*}%
for $f\in
\mathbb{C}
\left( T\right) $. Let us denote by $m_{T}:\mathcal{H\rightarrow
\mathbb{C}
}\left( T\right) $ the matrix coefficient $m_{T}\left[ v\right]
\left( g\right) =\left \langle v,\rho \left( g^{-1}\right) \phi
\right \rangle $ for $v\in \mathcal{H}$.

\noindent {\bf Lemma 1} (\cite{GH2})
\label{integral_lemma} {\em We have%
\begin{equation*}
\Theta _{T}=\mathcal{M}_{T}\circ m_{T}.
\end{equation*}
}


\subsubsection{Fast oscillator transforms}

In practice, it is desirable to have a "fast" algorithm for
computing the oscillator transform (FOT for short). We work in the
following setting. The
vector $v$ is considered in the standard realization $\mathcal{H=%
\mathbb{C}
}\left( \mathbb{F}_{p}\right) $ (see \ref{standard_subsub}), in
this context the oscillator transform gives the transition matrix
between the basis of delta functions and the basis $\left \{ \phi
_{\chi }\right \} $ of character vectors. We will show that when
$T$ is a split torus and for an
appropriate choice of $\phi $, the oscillator transform can be computed in $%
O\left( p\log \left( p\right) \right) $ arithmetic operations.
Principally, what we will show is that the computation reduces to
an application of DFT followed by an application of the standard
Mellin transform, both transforms admit a fast algorithm
\cite{CT}.

Assume $T$ is a split torus. Since all split tori are conjugated
to one another, there exists, in particular, an element $s\in Sp$
conjugating $T$ with the standard diagonal torus $A$. In more
details, we have a
homomorphism of groups $Ad_{s}:T\rightarrow A$ sending $g\in T$ to $%
Ad_{s}\left( g\right) =sgs^{-1}\in A$. Dually, we have a homomorphism $%
Ad_{s}^{\vee }:A^{\vee }\rightarrow T^{\vee }$ between the
corresponding groups of characters.

The main idea is to relate the oscillator transform with respect
to $T$ with the oscillator transform with respect to $A$. The
relation is specified in the following simple lemma.

\noindent {\bf Lemma 2} (\cite{GH2})
\label{relation_lemma} {\em We have%
\begin{equation}
\left( Ad_{s}^{\vee }\right) ^{\ast }\circ \Theta _{T,\phi
}=\Theta _{A,\rho \left( s\right) \phi }\circ \rho \left( s\right)
. \label{relation_eq}
\end{equation}
}


\noindent {\bf Remark 2} Roughly speaking, (\ref{relation_eq})
means that (up to a "reparametrization" of $T^{\vee }$ by $A^{\vee
}$ using $Ad_{s}^{\vee }$) the oscillator transform$\ $of a vector
$v\in \mathcal{H}$ with respect to the torus $T$ is the same as
the oscillator transform of the vector $\rho \left( s\right) v$ $\
$with respect to the diagonal torus $A$.


In order to finish the construction we need to specify two basic
facts about \ the Weil representation in the standard realization.

\begin{itemize}
\item The standard torus $A$ acts by (normalized) scalings, the precise
formula of $\rho \left( g\right) $ for $g=%
\begin{pmatrix}
a & 0 \\
0 & a^{-1}%
\end{pmatrix}%
\in A$ is%
\begin{equation*}
\rho \left( g\right)[f]\left( x\right) =\sigma \left( a\right)
f\left( ax\right) .
\end{equation*}

\item Every operator $\rho \left( g\right) ,$ $g\in Sp,$ can be written in
the form $\rho \left( g\right) =M_{g_{1}}\circ F\circ
M_{g_{2}}\circ S_{a}$ where $M_{g_{1}},M_{g_{2}}$ are the
operators of multiplication by some functions $g_{1},g_{2}\in
\mathbb{C}
\left( \mathbb{F}_{p}\right) $, $S_{a}$ is the operator of scaling
by $a\in \mathbb{F}_{p}^{\times }$ and $F$ is the DFT.

\begin{equation*}
F\left[ f\right] \left( y\right) =\frac{1}{\sqrt{p}}\sum
\limits_{x\in \mathbb{F}_{p}}\psi \left( yx\right) f\left(
x\right) .
\end{equation*}
\end{itemize}

Given a function $f\in
\mathbb{C}
\left( \mathbb{F}_{p}\right) $, \ applying formula (\ref{relation_eq}) with $%
\phi =\rho \left( s\right) ^{-1}\delta _{1}$ yields \
\begin{equation}
\Theta _{T,\phi }\left[ f\right] \left( Ad_{s}^{\vee }\left( \chi
\right) \right) =\frac{1}{p-1}\sum \limits_{a\in
\mathbb{F}_{p}^{\times }}\sigma \left( a\right) \overline{\chi
}\left( a\right)\rho \left( s\right) [f] \left( a\right) ,
\label{fast_eq}
\end{equation}%
for every $\chi \in A^{\vee }$. In conclusion, formula
(\ref{fast_eq}) implies that $\Theta _{T,\phi }\left[ f\right] $
can be computed by, first, applying the operator $\rho \left(
s\right) $ to $f$ and then applying Mellin transform to the
result.

\noindent {\bf Problem 1}

\label{question} {\em Does there exists a fast algorithm for
computing the oscillator transform associated to a
\textbf{non-split} torus? }

\subsection{Diagonalization of the discrete Fourier transform}

In this subsection we apply the previous development in order to
exhibit a canonical basis of eigenvectors for the DFT. We will
show that the DFT can
be naturally identified (up to a normalization scalar) with an operator $%
\rho \left( \mathrm{w}\right) $ in the Weil representation, where
\textrm{w}
is an element in a maximal torus $T_{\mathrm{w}}\subset Sp$. We take \textrm{%
w}$\in Sp=SL_{2}\left( \mathbb{F}_{p}\right) $ to be the Weyl
element
\begin{equation*}
\mathrm{w}=%
\begin{pmatrix}
0 & 1 \\
-1 & 0%
\end{pmatrix}%
.
\end{equation*}

\noindent {\bf Lemma 3}(\cite{GH2}) \label{DFT_lemma} {\em We have
\begin{equation*}
F=C\cdot \rho \left( \mathrm{w}\right) ,
\end{equation*}
where $C=i^{\frac{p-1}{2}}.$ }


Lemma 3 
implies that the diagonalization problems
of the operators $F$ and $\rho \left( \mathrm{w}\right) $ are
equivalent. The second problem can be approached using
representation theory, which is what we are going to do next.

Let us denote by $T_{\mathrm{w}}$ the centralizer of \textrm{w} in
$Sp$, namely $T_{\mathrm{w}}$ consists of all elements $g\in Sp$ such that $g$%
\textrm{w}$=$\textrm{w}$g$, in particular we have that \textrm{w}$\in T_{%
\mathrm{w}}$.

\noindent {\bf Proposition 1}
(\cite{GH2})\label{fourier_torus_prop} {\em The group
$T_{\mathrm{w}}$ is a maximal torus. Moreover the split type of
$T_{\mathrm{w}}$ depends on the prime $p$ in the
following way: $T_{\mathrm{w}}$ is a split torus when $p\equiv 1($mod $4)$ and is a non-split torus when $p\equiv 3 ($mod $4)$.}


Proposition 1 
has several consequences.
First consequence is that choosing a unit character vector $\phi
_{\chi }\in
\mathcal{H}_{\chi }$ for every $\chi \in \mathsf{Spec}_{T_{\mathrm{w}%
}}\left( \mathcal{H}\right) $ gives a canonical (up to normalizing
unitary
constants) choice of eigenvectors for the DFT \footnote{%
In the case $T_{w}$ is a split torus there is a slight ambiguity
in the choice of a character vector with respect to $\sigma
_{T_{w}}$. This ambiguity can be resolved by further investigation
which we will not discuss here.}. Second, more mysterious
consequence is that although the formula of the DFT is uniform in
$p$, its qualitative behavior changes dramatically between the
cases when $p\equiv 1($mod $4) $ and $p\equiv
3($mod $4)$. This is manifested in the structure
of the group of symmetries: In the first case, the group of
symmetries is a split torus consisting of $p-1$ elements and in
the second case it is a non-split torus consisting of $p+1$
elements. It also seems that the structure of the
symmetry group is important from the algorithmic perspective, in the case $%
p\equiv 1($mod $4)$ we built a fast algorithm for
computing $\Theta $, while in the case $p\equiv 3($mod $4)$ the existence of such an algorithm remains
open (see Problem 1). 

\subsubsection{ Multiplicities of eigenvalues of the DFT}

Considering the group $T_{\mathrm{w}}$ we can give a transparent
computation
of the eigenvalues multiplicities for the operator $\rho \left( \mathrm{w}%
\right) $. First we note that, since \textrm{w} is an element of
order $4$,
the eigenvalues of $\rho \left( \mathrm{w}\right) $ lies in the set $%
\left \{ \pm 1,\pm i\right \} $. For $\lambda \in \left \{ \pm
1,\pm i\right
\} $, let $m_{\lambda }$ denote the multiplicity of the eigenvalue $%
\lambda $. We observe that%
\begin{equation*}
m_{\lambda }=\bigoplus \limits_{\chi \in I_{\lambda }}\dim
\mathcal{H}_{\chi },
\end{equation*}%
where $I_{\lambda }$ consists of all characters $\chi \in \mathsf{Spec}_{T_{%
\mathrm{w}}}\left( \mathcal{H}\right) $ such that $\chi \left( \mathrm{w}%
\right) =\lambda $. The result now follows easily from Theorem 3, 
 applied to the torus $T_{\mathrm{w}}$. We treat separately the
split an non-split cases.

\begin{itemize}
\item Assume $T_{\mathrm{w}}$ is a split torus, which happens when $%
p=1($mod $4)$, namely, $p=4l+1$, $l\in
\mathbb{N}
$. Since $\dim \mathcal{H}_{\chi }=1$ for $\chi \neq \sigma _{T_{\mathrm{w}%
}} $ it follows that $m_{\pm i}=\frac{p-1}{4}=l$. We are left to
determine
the values of $m_{\pm 1}$, which depend on whether $\sigma _{T_{\mathrm{w}%
}}\left( \mathrm{w}\right) =$\textrm{w}$^{\frac{p-1}{2}}$ is $1$
or $-1$.

Since \textrm{w} is an element of order $4$ in $T_{\mathrm{w}}$ we
get that
\begin{equation*}
\sigma _{T_{\mathrm{w}}}\left( \mathrm{w}\right) =\left \{
\begin{array}{cc}
1, & p\equiv 1\left( \text{mod}\, 8\right) , \\
-1,\text{ \ } & \text{\ }p\equiv 5\left( \text{mod}\, 8\right) ,%
\end{array}%
\right.
\end{equation*}
which implies that when $p\equiv 1($mod $8) $ then $%
m_{1}=l+1 $ and $m_{-1}=l$ and when $p\equiv 5($mod $8)$ then $m_{1}=l $ and $m_{-1}=l+1$.

\item Assume $T_{\mathrm{w}}$ is a non-split torus, which happens when $%
p\equiv 3($mod $4) $, namely, $p=4l+3$, $l\in
\mathbb{N}
$. Since $\dim \mathcal{H}_{\chi }=1$ for $\chi \neq \sigma _{T_{\mathrm{w}%
}} $ it follows that $m_{\pm i}=\frac{p+1}{4}=l+1$. The values of
$m_{\pm 1}$
depend on whether $\sigma _{T_{\mathrm{w}}}\left( \mathrm{w}\right) =$%
\textrm{w}$^{\frac{p+1}{2}}$ is $1$ or $-1$. Since \textrm{w} is
an element of order $4$ in $T_{\mathrm{w}}$ we get that
\begin{equation*}
\sigma _{T_{\mathrm{w}}}\left( \mathrm{w}\right) =\left \{
\begin{array}{cc}
1, & p\equiv 7\left( \text{mod}\, 8\right) , \\
-1,\text{ \ } & p\equiv 3\left( \text{mod}\, 8\right) ,%
\end{array}%
\right.
\end{equation*}
which implies that when $p\equiv 7($mod $8) $ then
$m_{1}=l$
and $m_{-1}=l+1$ and when $p\equiv 3($mod $8) $ then $%
m_{1}=l+1$ and $m_{-1}=l$.
\end{itemize}

Summarizing, the multiplicities of the operator $\rho \left( \mathrm{w}%
\right) $ are
\begin{equation}
\begin{tabular}{|l|l|l|l|l|}
\hline & $m_{1}$ & $m_{-1}$ & $m_{i}$ & $m_{-i}$ \\ \hline
$p=8k+1$ & $2k+1$ & $2k$ & $2k$ & $2k$ \\ \hline $p=8k+3$ & $2k$ &
$2k+1$ & $2k+1$ & $2k+1$ \\ \hline $p=8k+5$ & $2k+1$ & $2k+2$ &
$2k+1$ & $2k+1$ \\ \hline $p=8k+7$ & $2k+2$ & $2k+1$ & $2k+2$ &
$2k+2$ \\ \hline
\end{tabular}
\label{table1}
\end{equation}

Considering now the DFT operator $F$. If we denote by $n_{\mu }$,
$\mu \in \left \{ \pm 1,\pm i\right \} $ the multiplicity of the
eigenvalue $\mu $ of $F $ then the values of $n_{\mu }$ can be
deduced from table \ref{table1} by
invoking the relation $n_{\mu }=m_{\lambda }$ where $\lambda =i^{\frac{p-1}{2%
}}\cdot \mu $ (see Lemma 3). 
Summarizing, the multiplicities of the DFT are
\begin{equation*}
\begin{tabular}{|l|l|l|l|l|}
\hline & $n_{1}$ & $n_{-1}$ & $n_{i}$ & $n_{-i}$ \\ \hline
$p=4l+1$ & $l+1$ & $l$ & $l$ & $l$ \\ \hline $p=4l+3$ & $l+1$ &
$l+1$ & $l+1$ & $l$ \\ \hline
\end{tabular}%
\end{equation*}
For a comprehensive treatment of the multiplicity problem from a
more classical point of view see \cite{AT}. \ Other applications of Lemma 3 
appear in \cite{GHH}.

\end{document}